\documentclass[twocolumn,floatfix,showpacs,amsmath,amssymb,pra]{revtex4}

\usepackage{dcolumn} 
\usepackage{amsmath,bbm}
\usepackage{amsfonts,amssymb}
\usepackage{graphicx}
\usepackage{epsfig}
\usepackage{times}
\usepackage{verbatim}

\newcommand{\be}{\begin{equation}}
\newcommand{\ee}{\end{equation}}
\newcommand{\bq}{\begin{eqnarray}}
\newcommand{\eq}{\end{eqnarray}}
\newcommand{\ba}{\begin{align}}
\newcommand{\ea}{\end{align}}

\newcommand{\dg}{\dagger}
\newcommand{\ket}[1]{ | \, #1 \rangle}
\newcommand{\bra}[1]{ \langle #1 \,  |}
\newcommand{\braket}[2]{\left\langle\, #1\,|\,#2\,\right\rangle}

\newcommand{\proj}[1]{\ket{#1}\bra{#1}}

\newcommand{\tr}[1]{{\rm tr}\left[{#1}\right]}

\newcommand{\Df}[1]{\frac{d}{d #1}}

\newcommand{\bC}{\mathbbm{C}}

\newcommand{\cL}{\mathcal{L}}

\newcommand{\cO}{\mathcal O}

\newcommand{\half}{\frac{1}{2}}

\def\qed{\leavevmode\unskip\penalty9999 \hbox{}\nobreak\hfill
     \quad\hbox{\leavevmode  \hbox to.77778em{%
               \hfil\vrule   \vbox to.675em%
               {\hrule width.6em\vfil\hrule}\vrule\hfil}}
     \par\vskip3pt}
    {\hspace*{\fill}$\Box$\vspace{1.5ex}\par}

\newcommand{\Sp}{\,\,\,\,\,\,}

\begin{document}

\title{Runtime of unstructured search with a faulty Hamiltonian oracle}
\author{Kristan Temme}
\affiliation{Center for Theoretical Physics, Massachusetts Institute of Technology, Cambridge, 02139 MA, USA and \\
	         IQIM, California Institute of Technology, Pasadena, CA 91125, USA}
\date{\today}

\begin{abstract} 
\noindent

We show that it is impossible to obtain a quantum speed-up for a faulty Hamiltonian oracle. The effect of dephasing noise to this continuous time oracle model has first been investigated in [{\em Phys. Rev. A}, {68}, {\bf 5}, {\em 052313}, {(2003)}]. The authors consider a  faulty oracle  described by a continuous time master equation that acts as dephasing noise in the basis determined by the marked item. The analysis focuses on the implementation with a particular driving Hamiltonian. A universal lower bound for this oracle model, which rules out a better performance with a different driving Hamiltonian has so far been lacking. Here, we derive an adversary type lower bound  which shows that the evolution time $T$ has to be at least in the order of $N$, i.e. the size of the search space, when the error rate of the oracle is constant. This means that quadratic quantum speed-up vanishes and the runtime assumes again the classical scaling. For the standard quantum oracle model this result was first proven in [{\em ICALP 2008}, Part I, {\em LNCS} {\bf 5125}, pp {\it 773 - 781} (2008)]. Here, we extend this result to the continuous time setting.  

\end{abstract}

\maketitle
\section{Introduction}
\label{sec:intro}

The Hamiltonian oracle \cite{Farhi} model can be seen as a continuous time analogue of the unstructured search problem, which is also known as Grover's problem. 
In unstructured search, the task is to find one marked item, which is commonly labeled by $w$, out of $N$ possible items. It is known, that on a classical computer on average at least ${\cal O}(N)$ queries to the oracle are needed to find the marked item. One of the major breakthroughs in search of quantum algorithms was that this bound could be beaten on a quantum computer. Lov Grover showed that a quantum algorithm exists, which only queries the oracle ${\cal O}(\sqrt{N})$ times \cite{Grover}. It can be shown that this quadratic speed-up is optimal \cite{Benett}. Hence, no algorithm can outperform Grover's search for this problem.

However, the investigation of Grover's algorithm in the presence of noise \cite{Shenvi,obenland1999simulating,long2000dominant,gawron2012noise}, has 
shown that this quadratic speed-up is very fragile. In quantum query algorithms two classes of noise models have been considered. One class of noise model considers coherent errors \cite{buhrman2007robust,shapira2003effect}, whereas the other class models errors in terms of either dephasing or bit-flip errors 
\cite{Shenvi,obenland1999simulating,long2000dominant,gawron2012noise,Vrana}. For the latter class, the quadratic speed-up of Grover's algorithm vanishes and the runtime assumes a linear scaling \cite{Shenvi}.  Regev and Schiff have recently proven, that no other query algorithm that has access to a dephasing oracle can outperform this scaling \cite{Regev}. 

The effect of dephasing noise on the continuous time analogue of Grover's algorithm in the Hamiltonian oracle setting has also been investigated by Shenvi et al. \cite{Shenvi}. Similar to the discreet case, the authors have found that when considering a constant error rate the quadratic speed-up over the best classical solution vanishes. This was investigated by the direct analysis of a specific quantum algorithm subject to an appropriately chosen noise model. The question that remained open is, whether this performance in the continuous time (Hamiltonian oracle) algorithm found by the authors is in fact optimal, i.e no other algorithm could perform better.  We will show, as could be expected, that this is indeed the case by extending the proof of Regev and Schiff to the Hamiltonian oracle setting.\\ 

In \cite{Shenvi}, Shenvi et al. considered the effect of phase fluctuations on the query term of the Hamiltonian model. The authors showed that such a fluctuating term leads to dephasing of the density matrix. Such an effect can best be described by a continuous time master equation of Lindblad form.  The most general form for such an equation is given by \cite{Lindblad}:
\be\label{master}
\Df{t}\rho = -i\left[H,\rho\right] + \sum_i L_i \rho L_i^\dg -\half\left\{L_i^\dg L_i, \rho \right\}_{+}.
\ee
Here, the Hamiltonian $H$  drives the coherent evolution, whereas the Lindblad operators $L_i$ can lead to loss of coherence and damping.

\section{Faulty Hamiltonian oracle model}
\label{sec:FaultyModel}
The general Hamiltonian oracle model can be described as follows: Rather than applying a sequence of unitaries, as is done in the circuit model of quantum computation, one considers the evolution of some quantum state subject to the Schr\"odinger equation 
\be
	i\Df{t}\ket{\psi} = H(t)\ket{\psi}.
\ee
The computation is encoded in the Hamiltonian $H(t)$, which is allowed to vary in time. The search problem can be encoded in terms of a Hamiltonian oracle, which is a term present in the Hamiltonian. It is important to note, that even though we are allowed to choose particular Hamiltonians $H(t)$ that realize the quantum algorithm, we do not have control over the term which corresponds to the Hamiltonian oracle.

For the unstructured search problem we consider the Hilbert space $\bC^N$ where the basis states $\{\ket{k}\}_{k=1\ldots N}$ label the items in the search space of size $N$. The task in the unstructured search problem is to find a single marked item we label by $\ket{w}$, also referred to as the winner. The general goal is to  construct a Hamiltonian that drives the evolution towards the state $\ket{w}$. In general such a Hamiltonian is of the form 
\be\label{driver}
	H(t) = E \proj{w} + H_D(t).
\ee
Here the projector on the winner $H_w = E \proj{w}$ encodes the Hamiltonian oracle. The actual computation, we have control over, is encoded in the driving Hamiltonian $H_D(t)$. 

A particular driver that solves the unstructured search problem is given by the the projector on the superposition of all basis states in the search space. This corresponds to the choice $H_D = E \proj{s}$, where $\ket{s}$ is the superposition of all basis states given by the coherent 'mixture' $\ket{s} = N^{-\half} \sum_{k=1}^N \ket{k}$. The overlap between the winner and the mixture is given by $ \braket{w}{s} = N^{-\half}$. For such a driving Hamiltonian it was shown \cite{Farhi}, that the time to generate constant overlap with the winner starting from the coherent mixture scales as $t = {\cal O}(\sqrt{N} E^{-1})$.  Moreover, it was shown that no other choice of driver $H_D(t)$ can outperform this scaling.\\

The analysis of this problem assumes a perfect implementation of the oracle Hamiltonian $\proj{w}$. However, in a realistic application one would expect, that the oracle is subject to some form of noise. Let us assume that the magnitude of the oracle Hamiltonian is subject to small fluctuations. That is we assume that the oracle is of the form
\be
	H_w^\xi = \left(E + \xi(t) \right)  \proj{w},
\ee 
where $\xi(t)$ is a stochastic variable for which the Markov assumption holds. This variable satisfies $\int_0^\pi \xi(t) dt = \epsilon$, where $\epsilon$ is distributed according to a Gaussian distribution with variance $s$. As was shown in \cite{Shenvi}, such a fluctuating term in the oracle model leads to dephasing in the basis determined by the oracle with a rate $\Gamma = \frac{s^2}{2 \pi}$.  We therefore state the {\it noisy Hamiltonian oracle model} in terms of a dephasing master equation of the form (\ref{master}) with a single dephasing Lindblad operator $L_1 = \sqrt{\Gamma}\proj{w}$. The full noisy oracle model describes the evolution of a density matrix $\rho$ according to the equation
\be\label{oracle}
	\Df{t}\rho = -i \left[H(t),\rho \right] + \cL_w(\rho),
\ee 
where we denote 
\be
\cL_w(\rho) = \Gamma\left(\proj{w} \rho \proj{w} - \half\left\{\proj{w},\rho\right\}_+\right).
\ee
The coherent evolution is now given again by the error-free Hamiltonian $H(t) = H_w + H_D(t)$. \\

We will have to compare the evolution of the system where the oracle is present with {\it the evolution in absence of the oracle} to see how much progress is made towards achieving the goal, i.e. generating sufficient overlap with the target state $\proj{w}$. To this end we also state the evolution in absence of the oracle which is given by
\be
	\Df{t}\rho = -i\left[H_D(t),\rho\right].
\ee
Note, that since no oracle term $H_w$ is present, we assume that this evolution is not subject to noise and hence the system only evolves unitarily with the driver $H_D(t)$. Unlike the evolution subject to the noisy oracle, the evolution in the absence of an oracle retains the purity of a pure initial state.  

\section{Runtime lower bound}
\label{sec:bound} 
We now proceed to derive the lower bound on the runtime to find the marked state when we can make use of the noisy oracle.  

We find that the noisy oracle with a constant dephasing rate $\Gamma$ cannot yield a quantum speed up over the classical bound. We can state 
as our main result:\\

\textbf{Main Result:}
{\it Every driver Hamiltonian that finds the marked state $\ket{w}$ with probability $p > 2^{-1/2}$ has to evolve on avarage for a time $T$ at least}
\be\label{mainBound}
	T \geq N \frac{2\Gamma (2p^2 -1) }{\Gamma^2 + 4E^2}.
\ee\\

The strategy for showing this is the following: First we construct a progress measure which has to be larger than $\cO(N)$ after the evolution time $T$ of the algorithm. We then derive an upper bound on the growth rate of the progress measure. From this we can infer the bound on the runtime of the algorithm.  

We compare the evolution of the state that evolves according to the Hamiltonian oracle with respect to the evolution of the system where no oracle is present. In order to differ between the two cases, we need to define a progress measure. A suitable progress measure can be defined from the Frobenius norm difference between two states. We write 
\be
	F_t^w = \|\rho^w_t - \rho^0_t\|_F^2.
\ee
Recall that the Frobenius norm is defined as $\|A\|_F = \sqrt{\tr{A^\dagger A}}$. Since we are interested in the performance of the algorithm for an arbitrary marked item $\ket{w}$, we need to consider 
the (unnormalized) average over all marked items. We define the {\it progress measure} as
\be
	F_t = \sum_{w=1}^{N} F_t^w.
\ee

\paragraph{Lower bound to the progress measure:\\}
The  lower bound to the progress measure after time $T$ is obtained from the following argument: For the algorithm to be successful, we want to be able to find the state $\ket{w}$ at least with a fixed probability $p$ after the algorithm has completed.  To this end the trace norm difference between the state $\rho^w_T$ which has evolved for time $T$ subject to the oracle (\ref{oracle})  has to differ by 
\be
	\half \|\rho^w_T - \rho^0_T\|_{tr} \geq p
\ee 
from the state $\rho^0_T$ which evolved in absence of the oracle. The trace norm of some operator $A$ on the space $\bC^N$ is defined as $\|A\|_{tr} = \tr{\sqrt{A^\dg A}}$. This distance has an operational interpretation and indicates the best statistic distinguishability by quantum measurements between the two states \cite{Fuchs}.  Recall that the evolution without oracle preserves purity. We therefore know that $\rho^0_T = \proj{\varphi_T}$ if we started without loss of generality in some pure state $\proj{\varphi_0}$. A well known bound \cite{Fuchs} on the trace distance between two quantum states can be given in terms of the fidelity. We can therefore bound 
\bq\label{numBnd}
	p \leq \half \|\rho^w_T - \rho^0_T\|_{tr} \leq \sqrt{1 - \bra{\varphi_T}\rho^w_T\ket{\varphi_T}}.
\eq

Since we have that $\tr{\left(\rho^0_T\right)^2} = 1$ and $\tr{\left(\rho^w_T\right)^2} \geq 0$, we know that after some $T$ the value of  $F_T^w$ has to be 
\bq
F_T^w &=& \tr{\left(\rho^w_T\right)^2} + \tr{\left(\rho^0_T\right)^2} - 2 \bra{\varphi_T}\rho^w_T\ket{\varphi_T}\\\nonumber 
        &\geq& 1 - 2 \bra{\varphi_T}\rho^w_T\ket{\varphi_T} \geq (2p^2 -1). 
\eq
The final bound is obtained from (\ref{numBnd}). After summing over all marked items, the average progress measure is bounded by
\be\label{upper}
	F_T = \sum_{w=1}^N F_T^w \geq N(2p^2 -1).
\ee

\paragraph{The growth rate of the progress measure:\\}
We now have to see how long it will take for the evolution of the progress measure to reach this value and will compute a bound on the rate by which it increases.  So we compute
\bq
\Df{t}F_t^w &=& \Df{t}\left(\tr{\left(\rho^w_t\right)^2} + 1 - 2\tr{\rho^w_t \rho^0_t}\right)\\
&=& 2\left(\tr{ \cL_w(\rho^w_t) {\rho}^w_t  }  - \tr{\cL_w(\rho^w_t) \rho^0_t}\right. \label{bth} \\\nonumber 
&& \left.+i\tr{\left[H_w,\rho^w_t\right] \rho^0_t}\right).
\eq

Note that the dependence on the driver Hamiltonian $H_D(t)$ has vanished. This is due to the fact that evolution of the driver in the oracle model cancels with the evolution of $\rho^0_t$. The evolution equation for the density matrix $\rho^w_t$ depends only on the winner $\proj{w}$. The other relevant projector is given by the pure state  $\rho^0_t = \proj{\varphi_t}$ which has evolved in the absence of the oracle.

Let us for convenience first consider the two-dimensional subspace spanned by the non-orthogonal vectors $\ket{w},\ket{\varphi_t}$. We can introduce the two orthogonal basis vectors $\ket{w},\ket{w^\perp}$ that span the same space so that we can write
\be
	\ket{\varphi_t} = \braket{w}{\varphi_t}\; \ket{w} + \sqrt{1-|\braket{w}{\varphi_t}|^2 }\; \ket{w^\perp}.
\ee
We proceed to patch these states with some orthonormal basis supported only on the complement of this two-dimensional space. The resulting basis is $\{\ket{w},\; \ket{w^\perp},\; \ket{\tilde{3}}, \ldots, \ket{\tilde{N}}\}$. To simplify the notation we define  $f \equiv \braket{w}{\varphi_t} \sqrt{1-|\braket{w}{\varphi_t}|^2 }$.

We evaluate the contributions to the derivative of the progress measure $F^w_t$ in eqn. (\ref{bth}), in this basis.  We see that both terms that depend on $\rho^0_t$ and $\rho^w_t$ are given by
\bq
&&\tr{\cL_w(\rho^w_t) \rho^0_t}              = -\frac{\Gamma}{2}\left( \left[\rho^w_t\right]_{w,w^\perp} \overline{f}  + \overline{\left[\rho^w_t\right]}_{w,w^\perp} f \right), \Sp \\
&&\tr{\left[H_w,\rho^w_t\right] \rho^0_t}  =  E\left( \left[\rho^w_t\right]_{w,w^\perp} \overline{f}  - \overline{\left[\rho^w_t\right]}_{w,w^\perp} f \right). \Sp 
\eq
Note that in these terms the only contribution from $\rho^w_t$ comes from the subspace spanend by $\ket{w},\ket{w^\perp}$. If we consider the remaining summand that only depends on $\rho^w_t$, we obtain 
\be
\tr{ \cL_w(\rho^w_t) {\rho}^w_t  }   =  -\Gamma |\left[\rho^w_t\right]_{w,w^\perp}|^2 - \Gamma \sum_{\tilde{k}=3}^N \left |\left[\rho^w_t\right]_{w,\tilde{k}} \right|^2.
\ee
Recall, that we want to find an upper bound on the evolution of the progress measure.  Therefore, we can only increase the bound on the progress measure by assuming that the state $\rho^w_t$ is only supported on the two dimensional subspace and therefore set  $\sum_{\tilde{k}=3}^N \left |\left[\rho^w_t\right]_{w,\tilde{k}} \right|^2 = 0$.  Now the derivative of $F_t^w$ only depends on the single matrix  element $x \equiv \left[\rho^w_t\right]_{w,w^\perp}$ \\

With the variables $x$ and $f$ defined earlier we can write for eqn. (\ref{bth}) 
\be\label{function}
\Df{t}F_t^w = 2\left( -\Gamma |x|^2 + \left(\frac{\Gamma}{2} +i E \right)x f^*  +  \left(\frac{\Gamma}{2} -i E \right) x^* f \right).
\ee
It is easy to see, that the RHS of (\ref{function}) becomes maximal for the choice $x_{opt} = \left(\frac{1}{2} - i\frac{E}{\Gamma} \right)f$. Furthermore, since $|f|^2 \leq |\braket{w}{\varphi_t}|^2$ we can state the inequality
\be
	\Df{t}F_t^w \leq \frac{\Gamma^2 +4E^2}{2 \Gamma}	|\braket{w}{\varphi_t}|^2.
\ee
Note, that $\ket{\varphi_t}$ is a normalized state. We therefore have, that the sum over all winners is bounded by 
\be\label{inequ_funky}
	\Df{t}F_t = \sum_{w=1}^N \Df{t} F_t^w \leq \frac{\Gamma^2 +4E^2}{2 \Gamma}
\ee
Integrating inequality (\ref{inequ_funky}) with the initial condition $F_0 = 0$, we find that 
\be
F_T \leq  \frac{\Gamma^2 + 4E^2}{2 \Gamma} T.
\ee 
Together with inequality (\ref{upper}), this leads to the bound on the minimal evolution time $T$ as stated in the main result (\ref{mainBound}).\\ 

When considering a fixed error rate $\Gamma$, we observe that the previous square root scaling in the database has now been reduced to a liner scaling, which is also what happens for the standard oracle model of quantum computation. The authors of \cite{Shenvi}, also considered what happens when one allows for an error rate that decreases in the size of the data base, i.e. $\Gamma = \alpha N^{-2\delta}$, where both $\alpha$ and $\delta$ are positive constants. With this error rate the runtime of the noisy Grover algorithm scales as $T = \cO(N^{1-2\delta})$, as long as $\delta \leq 1/4$. The bound in the main result reproduces the exact same scaling of the runtime. However, for $\delta  > 1/4$ the actual bound of the coherent evolution $T = \cO(N^{1/2})$ has to be considered since the bound given for the noisy oracle seizes to be tight. 

\section{Conclusions}
\label{sec:Concl} 
In conclusion, we have recovered the bound on the runtime of unstructured search which holds for the standard noisy oracle model also in the noisy Hamiltonian oracle model framework. With a constant dephasing error rate the quantum speed up breaks down and reduces to the known classical result of unstructured search. The techniques used here are very much in the spirit of the original proof \cite{Farhi} of the noise free Hamiltonian oracle model. The major difference is the noisy evolution described by the dephasing master equation and a new progress function, which uses the Hilbert-Schmidt norm between two density matrices, as apposed to the standard $L_2$-norm between two pure states. 

\section{Acknowledgments}
I would like to thank Edward Farhi for suggesting this research project to me. Moreover, I would like to thank Shelby Kimmel and Fernando Pastawski for insightful discussions. The author gratefully acknowledges the support from the Erwin Schr\"odinger fellowship, Austrian Science Fund (FWF): J 3219-N16. This work was supported by the Institute for Quantum Information and Matter, a NSF Physics Frontiers Center with support of the Gordon and Betty Moore Foundation (Grants No. PHY-0803371 and PHY-1125565).

\end{document}